\documentclass[prd,10pt,aps,twocolumn,notitlepage,amsmath,nofootinbib,superscriptaddress,altaffilletter,showpacs,showkeys]{revtex4-2}

\usepackage{graphicx}
\usepackage[svgnames,table]{xcolor}
\usepackage{hyperref}
\hypersetup{
    colorlinks=true,
    linkcolor=red,     
    citecolor=blue,     
    urlcolor=red       
}
\usepackage[english]{babel}

\newcommand*{\arraycolor}[1]{\protect\leavevmode\color{#1}}
\newcolumntype{A}{>{\columncolor{blue!20!white}\centering\arraybackslash}m{1.5cm}} 
\newcolumntype{B}{>{\columncolor{blue!20!white}\centering\arraybackslash}m{3.5cm}} 
\newcolumntype{C}{>{\columncolor{blue!20!white}\centering\arraybackslash}m{3.5cm}} 
\newcolumntype{D}{>{\columncolor{blue!20!white}\centering\arraybackslash}m{3.5cm}} 
\begin{document}

\title{Stability of the spacetime  of a magnetized compact object}
\author{Eveling C. Ribeiro}\email{evelingmilena@usp.br}
\author{L. Formigari}\email{lucas.f.formigari@gmail.com}
\author{Marcos R. Ribeiro Jr.}\email{marcosribeiro@usp.br}
\affiliation{
Instituto de F\'\i sica, Universidade de S\~ao Paulo,  05508-09 S\~ao Paulo, SP,  Brazil}

\author{Elcio Abdalla}\email{eabdalla@usp.br}
\affiliation{
Instituto de F\'\i sica, Universidade de S\~ao Paulo,  05508-09 S\~ao Paulo, SP,  Brazil}
\affiliation{Universidade Estadual da Paraíba,
58429-500 Campina Grande, PB, Brazil}
\affiliation{Departamento de Física,
Universidade Federal da Paraíba, 58059-970  João Pessoa, PB, Brazil}

\author{Bertha Cuadros-Melgar}\email{bertha@usp.br}
\affiliation{Escola de Engenharia de Lorena, Universidade de S\~ao Paulo, 
  12602-810 Lorena, SP, Brazil}

\author{C. Molina}\email{cmolina@usp.br}
\affiliation{ Escola de Artes, Ci\^encias e Humanidades, Universidade de S\~ao Paulo,
  03828-000, S\~ao Paulo, SP, Brazil}

\author{Amilcar R. de Queiroz}\email{amilcarq@gmail.com  }
\affiliation{Unidade Acad\^emica de F\'\i sica, Universidade Federal de Campina Grande, 58429-900  Campina Grande,  
PB, Brazil}
\author{Alberto Saa}\email{asaa@ime.unicamp.br}
\affiliation{Departamento de Matem\'atica Aplicada, Universidade Estadual de Campinas, 13083-859 Campinas, SP, Brazil}

\begin{abstract}
We investigate the stability of scalar perturbations around a magnetized stationary compact object in General Relativity. The considered object is one of the simplest exact solutions of Einstein electrovacuum equations corresponding to a spheroidal body endowed with a   dipole magnetic moment. It is effectively constructed by imposing a perfect reflection (mirror) boundary condition on a central region of the  Gutsunaev-Manko spacetime. A time-domain analysis of the perturbations reveals a quasinormal phase followed by a power-law   decaying tail.  Our findings suggest that the exterior region of the magnetized compact object is stable in the entire parameter space. Moreover, the system tends to become generically more stable the stronger the magnetization of the central object is.  Such findings can be useful for the qualitative understanding of more realistic astrophysical situations involving highly magnetized sources. 
\end{abstract}
\date{\today}
\maketitle

\section{Introduction}

Einstein field equations display an extraordinary diversity of solutions with different and exciting properties. 
The Schwarzschild metric {is} the simplest possible solution describing an astrophysical object, and many of its generalizations  have astrophysical relevance \cite{ PhysRev.56.455, 1642270, stephani2009exact}.  From an observational point of view, its most reasonable stationary generalizations should include rotation and magnetization since these are typical physical attributes of astrophysical sources.

We are mainly concerned here with the case of magnetized axisymmetric non-rotating bodies as described by the exterior region of the Gutsunaev-Manko spacetime \cite{gutsunaev1987gravitational,gutsunaev1988family,manko1989new,manko1989}, which can be asymptotically interpreted as a spherically  symmetric core endowed with a dipole magnetic moment. Such a solution is certainly of astrophysical interest since it provides a simple yet viable description of the exterior region of 
{  non-rotating magnetized bodies.
The closest magnetized astrophysical body to the Earth is the Sun. Its field strength  is rather feeble by astrophysical standards, approximately $1$ G on its surface, only twice the Earth's magnetic field strength, but its dipole component gives origin to the 
heliospheric magnetic field which is responsible for  many relevant   phenomena in the solar system \cite{TheSun}. Nevertheless, it is well known that the Sun has strong local perturbations on its magnetic field, reaching strengths of the order of $10^3$ G on some spots. Moreover, the Sun dipole is known to exhibit  a regular 
dynamics, changing its polarity in approximate  11-year periods. Although our model might be useful to describe
fast phenomena -- compared with the dipole changes period -- of 
 low magnetization 
stars like the Sun, we think it is particularly more interesting for the study of   
magnetars, {\it i.e.},  highly magnetized neutron stars, typically with low angular momentum  \cite{Kaspi_2017}. This class of solutions has been gaining prominence in the recent literature \cite{deFreitas:2017rbk, Polanco:2023jsp}. The intense magnetic fields (of the order of $10^{13}$ to $10^{15}$ G) are prone to produce high-energy electromagnetic radiation phenomena {such as} bursts and flares of X-rays, gamma-rays, and perhaps even Fast Radio Bursts (FRB)  \cite{Kirsten_2020, zhang2020physical}. This possibility has gained interest due to the recently reported detection  \cite{ChimeFrb_1}  of a repeating FRB  from the magnetar SGR 1935+2154, which lies on the outskirts of our own Milky Way. The investigation of these phenomena is one of the main scientific goals of the soon-to-start operating BINGO radio telescope \cite{Abdalla_2022, santos2023bingo}. Previous enterprises concerning instability of localized solutions turn out to provide hints into modifications of interactions \cite{abdalla2019instability} or even traditional solutions with chosen specific perturbations \cite{Zhu:2014sya}.}

The exterior region of a magnetar is expected to be well described by an asymptotically flat electrovacuum spacetime with a magnetic field originating in the strong magnetization of the central core. This is precisely the situation corresponding to the Gutsunaev-Manko spacetime, whose stability we analyze here.  It is also possible to introduce a low angular momentum  into this picture through an approximate solution, see  \cite{morachaverri2024classical}. However, we  {opt to} consider only the non-rotating case and explore the effects of magnetization on the stability of the exterior region of these compact objects.

 {In the present work,} we adopt the simplest model of a compact object, namely a hard central core, which is assumed to be large enough to avoid the appearance of the   well-known singularities of the Gutsunaev-Manko spacetime. Hence, the exterior region of our compact object is, by construction, a regular asymptotically flat magnetized spacetime, and we will consider its stability by exploring a quasinormal mode (QNM) \cite{kokkotas1999quasi,konoplya2011quasinormal,abdalla2019instability} analysis of scalar perturbations with a total reflection (mirror) boundary condition on the surface of the central core. 
{ Such a total reflection at the body surface is not a new boundary condition, it has  indeed been  used before, see,  for instance,
\cite{fiziev2006exact}, where it was considered in the spherically symmetric (Schwarzschild) case. } 
 On physical grounds, one expects that such totally-reflected QNMs are stable, and we found that they indeed are. But, interestingly, we found that the presence of magnetization improves the stability of the spacetime, {\it i.e.}, the larger the magnetization, the larger the perturbations exponential suppression, and consequently the faster the relaxation of the system.  

This paper is organized as follows.  In Section 2 we will briefly review the construction of our compact object spacetime starting from the Gutsunaev-Manko solution. Section 3 will be devoted to the numerical scheme for the  {time-domain} calculation   of the scalar  perturbation. Features such as tails and quasi-normal modes are considered. The limit of small magnetization will be compared with an analytical calculation in the Schwarzschild limit. Finally, in Section 4 we will discuss our results and some of its astrophysical implications,  and suggest possible future directions for these studies.

\section{Magnetized compact objects}

The objects considered in this paper are assumed to be modeled by the simplest exact solutions of Einstein electrovacuum equations corresponding to a prolate  spheroidal body endowed with a dipole magnetic moment.  Effectively, they are constructed by imposing a perfect reflection (mirror) boundary condition on a central region of the Gutsunaev-Manko spacetime, whose main properties we briefly review  {in this section}.
For simplicity, we will also use the name ``star'' for the compact object hereafter.

The Gutsunaev-Manko solution is an  axisymmetric stationary metric which 
can be derived from the Ernst equations in the Weyl gauge \cite{manko1989, manko1993}. It is more conveniently expressed using the prolate spheroidal coordinates $(x,y,\varphi)$, which  {are} related to the usual cylindrical coordinates $\rho$ and $z$ by
 \begin{eqnarray}
   x &=& \frac{1}{2k}\left[\sqrt{\rho^2 + (z+k)^2} + \sqrt{\rho^2 + (z-k)^2}\right]  ,  \\
     y &=& \frac{1}{2k}\left[ \sqrt{\rho^2 + (z+k)^2} - \sqrt{\rho^2 + (z-k)^2} \right]   ,   
 \end{eqnarray}
 where $k$ is a dimensional constant related to the mass of the central body. The  {ranges} of the prolate coordinates $x$ and $y$, which are dimensionless by construction, are $x\ge 1$ and $-1\le y \le 1$. The inverse transformations are also useful 
\begin{eqnarray}
 \rho &=& k \sqrt{(x^2 - 1)(1 - y^2)} \, , \\ 
  z &=& kxy \, .
\end{eqnarray} 
The Gutsunaev-Manko metric reads
 \begin{widetext}
\begin{equation}
\label{eq:MAS:WeylGaugeGM}
    \mathrm{d}s^2 = -\left(\frac{x-1}{x+1} \right)f^2 \mathrm{d}\tau^2 +     \left(\frac{kg}{f}\right)^2 \left[
    \frac{x+1}{x-1}
    {\mathrm{d}x^2}  + \frac{(x+1)^2}{1-y^2}\mathrm{d}y^2 \right] + \frac{(x+1)^2(1-y^2)}{f^2}k^2\mathrm{d}\phi^2,
\end{equation}
\end{widetext}
where 
\begin{equation}
\label{eq:MAS:f_deff}
f = \frac{\left[x^2-y^2+\alpha^2(x^2-1)\right]^2 + 4\alpha^2x^2(1-y^2)}
{\left[ x^2-y^2+\alpha^2(x-1)^2 \right]^2 - 4\alpha^2y^2(x^2-1)} ,
\end{equation}
and 
\begin{equation}
\label{eq:MAS:g_deff}
g = \left\{\frac{ \left[x^2-y^2+\alpha^2(x^2-1) \right]^2 +   4\alpha^2x^2(1-y^2) }
{(\alpha^2+1)^2(x^2-y^2)^2}\right\}^2,
\end{equation}
with $\alpha$ being a dimensionless parameter, whose interpretation will become clear later. For $\alpha=0$, we have
$f=g=1$ and the Gutsunaev-Manko metric~(\ref{eq:MAS:WeylGaugeGM}) reduces to the Schwarzschild solution in prolate coordinates, with $x=1$ and  $-1\le y \le 1$ representing the event horizon. Also, the metric~(\ref{eq:MAS:WeylGaugeGM}),  with the functions~(\ref{eq:MAS:f_deff}) and (\ref{eq:MAS:g_deff}), is asymptotically flat. In the limit $x\to \infty$,
we have 
\begin{equation}
\label{asympt}
g_{\tau\tau}= -1 + \frac{2-6\alpha^2}{\alpha^2 + 1}x^{-1} + O(x^{-2}).
\end{equation}
Taking into consideration that for $x\to \infty$ we have $r=kx$ in spherical coordinates, the asymptotic expansion (\ref{asympt})
implies that Eq.~(\ref{eq:MAS:WeylGaugeGM}) corresponds to a central object with total mass
\begin{equation}
\label{eq:MAS:GMmass}
        M = \frac{  1 - 3 \alpha^2 }{1+ \alpha^2}k \, .
\end{equation}
{Expression~(\ref{eq:MAS:GMmass}) shows} that the physically reasonable solutions  {require} $3\alpha^2 < 1$.

The vector potential associated with the metric~(\ref{eq:MAS:WeylGaugeGM}) is $A_\mu = (0,0,0,A_\phi)$, where
\begin{eqnarray}
    A_\phi = && \frac{4k\alpha^3}{1+\alpha^2}(1-y^2)\times \\
    && \frac{  2(1+ \alpha^2)x^3 + (1 - 3 \alpha^2)x^2 + y^2 + \alpha^2}{ ( x^2 - y^2 + \alpha^2 (x^2 - 1))^2 + 4 \alpha^2 x^2 (1-y^2)  }\nonumber\label{Aphi},
\end{eqnarray}
{whose} asymptotic expansion for $x\to \infty $ is given by 
\begin{equation}
    \label{eq:MAS:Maglim}
    A_\phi =   \frac{8k^2 \alpha^3}{(1 + \alpha^2)^2 } \frac{\sin^2\theta }{r} + O(x^{-2}),
\end{equation}
with the spherical polar angle expressed as 
 as  $y=\cos\theta$ for $x\to \infty$.  From this asymptotic expansion,
 we have
 \begin{equation}
  \label{eq:MAS:Maglim2}
    A_bA^b = g^{\phi\phi}A_\phi A_\phi = \mu^2 \frac{\sin^2\theta}{r^4},
 \end{equation}
where 
    \begin{equation}
        \label{eq:MAS:GMdipole}
        \mu = \frac{8 k^2 \alpha^3}{(1 + \alpha^2)^2} \, .
    \end{equation}
    It is clear from (\ref{eq:MAS:Maglim2}) that 
 the vector potential~(\ref{Aphi}) corresponds 
to a central magnetic dipole with intensity given by (\ref{eq:MAS:GMdipole}).

Notice that with $\alpha$ in the range $-\frac{1}{\sqrt{3}}<\alpha < \frac{1}{\sqrt{3}}$ one can effectively describe all possibilities of mass $M$ and magnetization $\mu$ for the central object since
\begin{equation}
\label{fraction}
\frac{8\alpha^3}{(1-3\alpha^2)^2}= \frac{\mu}{M^2} \, .
\end{equation}

The Gutsunaev-Manko metric is known to be plagued with curvature singularities. For  {the} sake of illustration, let us consider the determinant of the metric~(\ref{eq:MAS:WeylGaugeGM})  on $x=1$. We have
\begin{equation}
\sqrt{|g_{ab}|} = \frac{4k^3}{\left(1+\alpha^2\right)^8} \left(1+\frac{4\alpha^2}{1-y^2}\right)^2 \, ,
\end{equation}
which clearly diverges at the poles $y = \pm 1$ for $\alpha\ne 0$, strongly suggesting the presence of a curvature singularity. 
In fact, the denominator of the function $f$ in Eq.~(\ref{eq:MAS:f_deff}) has zeros near $x=1$ and $y = \pm 1$,  indicating once more the existence of curvature singularities. To   avoid these problems, we will consider the Gutsunaev-Manko metric only for $x>x_{\mathrm{min}}$, with $x_{\mathrm{min}}$ large enough to hide the spacetime singularities. We interpret the surface $x=x_{\mathrm{min}}$ as the  boundary of a magnetized prolate spheroidal body.

\section{Stability of the exterior region}

\subsection{Scalar perturbations}

Our main goal here is to investigate the stability of the exterior region of the  magnetized  compact object introduced in the last section. For sake of simplicity, we will consider massless scalar perturbations, whose dynamics is  {described} by the standard  {Klein-Gordon} wave equation
\begin{equation}
\frac{1}{\sqrt{-g}}\partial_{a}\left(  \sqrt{-g}g^{ab
}\partial_{b}\Psi\right)  =0 \, .\label{dleq}
\end{equation}
It is more convenient for our purposes here to introduce the dimensionless time variable $t = k^{-1}\tau$. In terms of this new variable, the Klein-Gordon equation for the metric~(\ref{eq:MAS:WeylGaugeGM}) reads
\begin{eqnarray}
\label{wave}
 -   \partial_t^2\Psi + && \frac{(x-1)f^4}{(x+1)^3 g^2}\left[
 \partial_x (x^2-1) \partial_x\Psi  + 
 \partial_y (1-y^2) \partial_y\Psi \phantom{\frac{1}{ 2 }}\right. \nonumber \\
 &&\quad\quad\quad\quad \quad\quad\quad\quad \quad   +\left. \frac{g^2}{ 1-y^2 }\partial_\phi^2\Psi\right] = 0 \, .
\end{eqnarray}
This is our main equation, which will be studied with the total reflection boundary condition $\Psi=0$ on the compact body surface $x=x_{\mathrm{min}}$. 
{As we model the compact object as hard central core, we impose Dirichlet boundary conditions for the wave equation~(\ref{dleq}) with.}
Since the metric~(\ref{eq:MAS:WeylGaugeGM}) is asymptotically flat, one expects on physical grounds that any perturbation on the
exterior region of our compact body escapes toward infinity. However, as we will see, the magnetization of the central body affects
the  {perturbative} dynamics, making the relaxation faster. 

 {Let} us introduce the partial decomposition in spherical harmonics 
\begin{equation}
\label{deco}
\Psi(t,x,y,\phi) =\sum_{\ell, m}\frac{u_{\ell m}( t,x ) }{x+1}Y_{\ell}^{m}\left( y,\phi
\right) \, ,
\end{equation}
where we  employ  the following orthogonality condition
\begin{eqnarray}
\label{ortho} 
\left\langle Y_\ell^m ,Y_{\ell'}^{m'}  \right\rangle &=&
\int_{-1}^1 \int_0^{2\pi} Y_\ell^m(y,\phi) \bar Y_{\ell'}^{m'}(y,\phi) \mathrm{d}\phi \mathrm{d}y \nonumber \\ &=& \delta_{\ell \ell'}
\delta_{mm'},
\end{eqnarray}
and the tortoise coordinate $\tilde{x} = x + 2  \ln(x-1)$, which leads to
\begin{equation}
\partial_{\tilde x} =  \frac{x-1}{x+1} \partial_x,\label{jabuti}
\end{equation}
rendering  Eq.~(\ref{wave}) as
\begin{widetext}
\begin{equation}
\label{eq:MP:ScalWave3}
 - \sum_{\ell, m} Y_\ell^m  \partial_t^2u_{\ell m} + 
 \sum_{\ell, m} \frac{f^4}{ g^2} 
 \left\{ 
Y_\ell^m  \partial_{\tilde x}^2  u_{\ell m} - 
 \frac{x-1}{(x+1)^3} \left[\ell(\ell + 1) +
 \frac{m^2(g^2-1)}{   1-y^2  }+ \frac{2}{x+1} \right] Y_\ell^m  u_{\ell m} \right\} = 0.
\end{equation}
\end{widetext}
We now follow the usual steps of spectral methods. Multiply Eq.~(\ref{eq:MP:ScalWave3})  by $\bar Y^{m'}_{\ell'}$,  integrate  with respect to $y$ and $\phi$,
and invoke the orthogonality condition~(\ref{ortho}), leading to 
\begin{equation}
\label{eq:MP:ScalWave4}
 -    \partial_t^2u_{\ell m} + \sum_{\ell'}A_{\ell}^{\ell'  }
 \partial_{\tilde x}^2   u_{\ell' m } =  \sum_{\ell'}B_{\ell}^{\ell'}u_{\ell' m }
,
\end{equation}
where 
\begin{equation}
 \label{eq:MP:Amatrix}
 A_{\ell}^{\ell'  } =  \left\langle  Y_\ell^m ,\frac{f^4}{g^2}Y_{\ell'}^{m} \right\rangle   , 
 \end{equation}
and  
\begin{equation}
 \label{eq:MP:Bmatrix}
 B_{\ell}^{\ell' } = \frac{x-1}{(x+1)^3} \left( A_\ell^{\ell'}V_{\mathrm{eff}} +   {m^2}  \left\langle  Y_\ell^m ,\frac{g^2-1}{1-y^2} \frac{f^4}{g^2}Y_{\ell'}^{m} \right\rangle \right) \, ,
 \end{equation}
 with
 \begin{equation}
    \label{eq:MP:VeffScal}
    V_{\textrm{eff}} =  \ell(\ell + 1) +
    \frac{2}{x+1}     .
\end{equation}
It is  worth noting that for $g$ as described in Eq.~(\ref{eq:MAS:g_deff}), the limit
\begin{equation}
    \lim_{y \to \pm 1} \frac{g^2 - 1}{1 - y^2} = 1
\end{equation}
is well defined and, consequently, the integration performed in equations~(\ref{eq:MP:Bmatrix}) converges for every combination of $\ell$ and $\ell'$ in the external region. {It is also interesting
to consider the Schwarzschild limit $\alpha\to 0$ to grasp the underlying physical content of the equations
(\ref{eq:MP:ScalWave4}), (\ref{eq:MP:Amatrix}), (\ref{eq:MP:Bmatrix}), and (\ref{eq:MP:VeffScal}). In such a limit, we have
$f=g=1$ and, consequently, $A_{\ell}^{\ell'  } = \delta_{\ell}^{\ell'  }$. The $B_{\ell}^{\ell' }$ matrix (\ref{eq:MP:Bmatrix}) 
is diagonal in this case and equation (\ref{eq:MP:ScalWave4})   will read simply 
\begin{equation}
-    \partial_t^2u_{\ell m} + 
 \partial_{\tilde x}^2   u_{\ell m } =  \tilde{V}_{\rm eff} u_{\ell m },
\end{equation}
where
\begin{equation}
    \tilde V_{\textrm{eff}} =  \frac{x-1}{(x+1)^3}\left[ \ell(\ell + 1) +
    \frac{2}{x+1}  \right]  ,
\end{equation}
which is the usual effective potential for scalar (spin $0$) fields in the Schwarzschild case, as one can see by taking the limit of large
$ x \approx r$. It is quite natural to expect that for perturbations with other spins $s$, we end up with the usual effective potential of the type
\begin{equation}
    \tilde V_{\textrm{eff}}  \approx   \frac{ \ell(\ell + 1)}{r^2} +
    \frac{2(1-s^2)}{r^3}     ,
\end{equation}
for large $r$, but the expression for the $B_{\ell}^{\ell' }$ matrix (\ref{eq:MP:Bmatrix}) are much more intricate for
$s\ne 0$, rendering the analysis of other perturbation types 
in prolate   coordinates
 a quite laborious task. 
}

\begin{figure}[t]
        \includegraphics[width=0.9\linewidth]{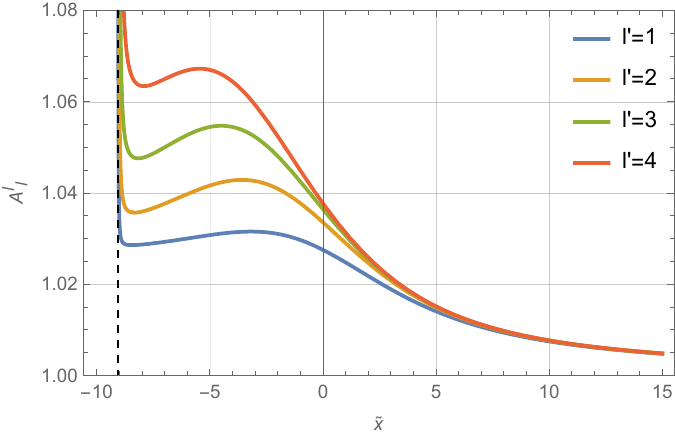}
        \includegraphics[width=0.9\linewidth]{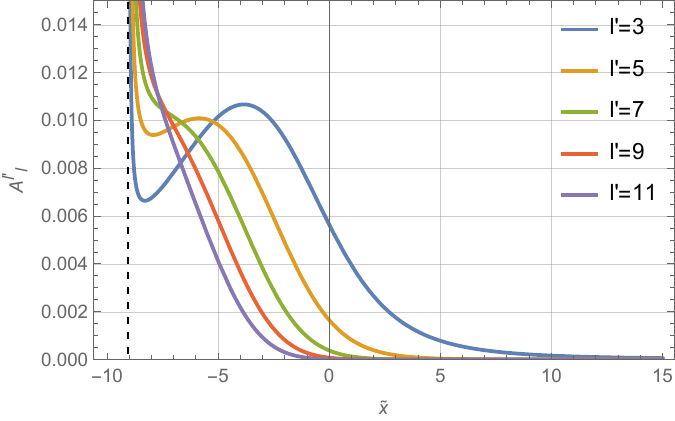}
    \caption{Top: $A_{\ell}^{\ell}$ diagonal terms. Bottom: $A_{\ell}^{\ell'}$ off diagonal terms. Both present divergences at $\tilde{x}_{\mathrm{sing}}\approx-9.07$ (dashed line) for $\alpha=0.057$.}\label{Amatrix}
\end{figure}

\begin{figure}[t]
        \includegraphics[width=0.9\linewidth]{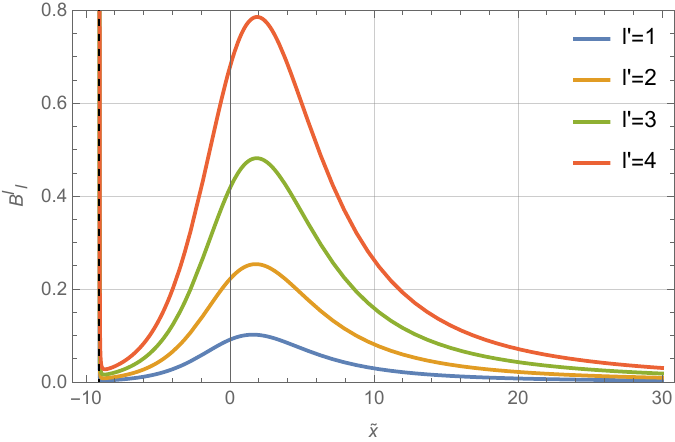}
        \includegraphics[width=0.9\linewidth]{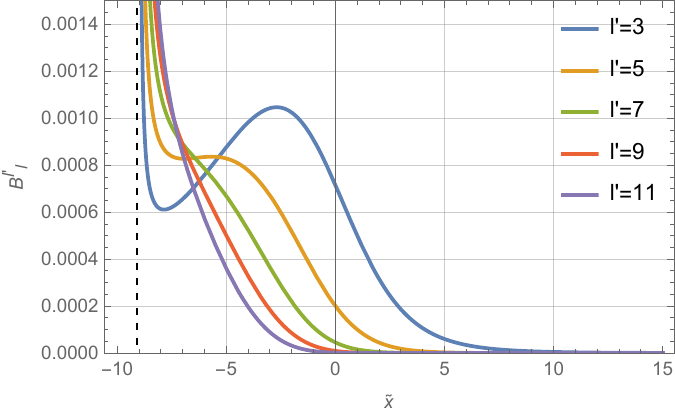}
    \caption{Top: $B_{\ell}^{\ell'}$ diagonal terms. Bottom: $B_{\ell}^{\ell'}$ off diagonal terms. Just as previous matrices, This one presents divergences at $\tilde{x}_{\mathrm{sing}}\approx-9.07$ (dashed line) for $\alpha=0.057$.}\label{Bmatrix}
\end{figure}

Equations~(\ref{eq:MP:ScalWave4}) form a set of coupled wave equations in the coordinates $(t,\tilde x)$. They are the basis of our numerical time-domain analysis, which main results are described in the following subsection.
Notice that, for the actual numerical calculations, one needs to truncate the sum in Eq. (\ref{eq:MP:ScalWave4}) by introducing a cutoff $\ell_{\mathrm{max}}$. As such, for a given a value of $m$, the parameter $\ell'$ takes integer values in the interval $[|m|, \ell_{\mathrm{max}}]$. The matrices couple only terms of the same parity, which stems from the integration over the $y$-coordinate of an even function when $\ell$ and $\ell'$ have the same parity, and odd when they have different parities.
The explicitly behavior of the matrices (\ref{eq:MP:Amatrix}) and (\ref{eq:MP:Bmatrix}) is displayed in Figures \ref{Amatrix} and \ref{Bmatrix}.
It is worth noticing that, provided $\ell_{\mathrm{max}}$ is large
enough, the QNM frequencies analysis is effectively  independent of
the employed cutoff, and this can be understood due to the tiny effects of the off-diagonal terms of the matrices $A^{\ell}_{\ell}$ and $B^{\ell}_{\ell}$, as long as $x_{\mathrm{min}}$ is far enough from singularities. Otherwise, the $\ell_{\mathrm{max}}$ might be relevant, especially for higher values of multipole momenta. In light of this, in all numerical calculations we ensures that $x_{\mathrm{min}}$ is sufficiently distant from singularities. 

\subsection{Numerical results}
\begin{figure}[bt]
    \centering
    \includegraphics[scale=0.5]{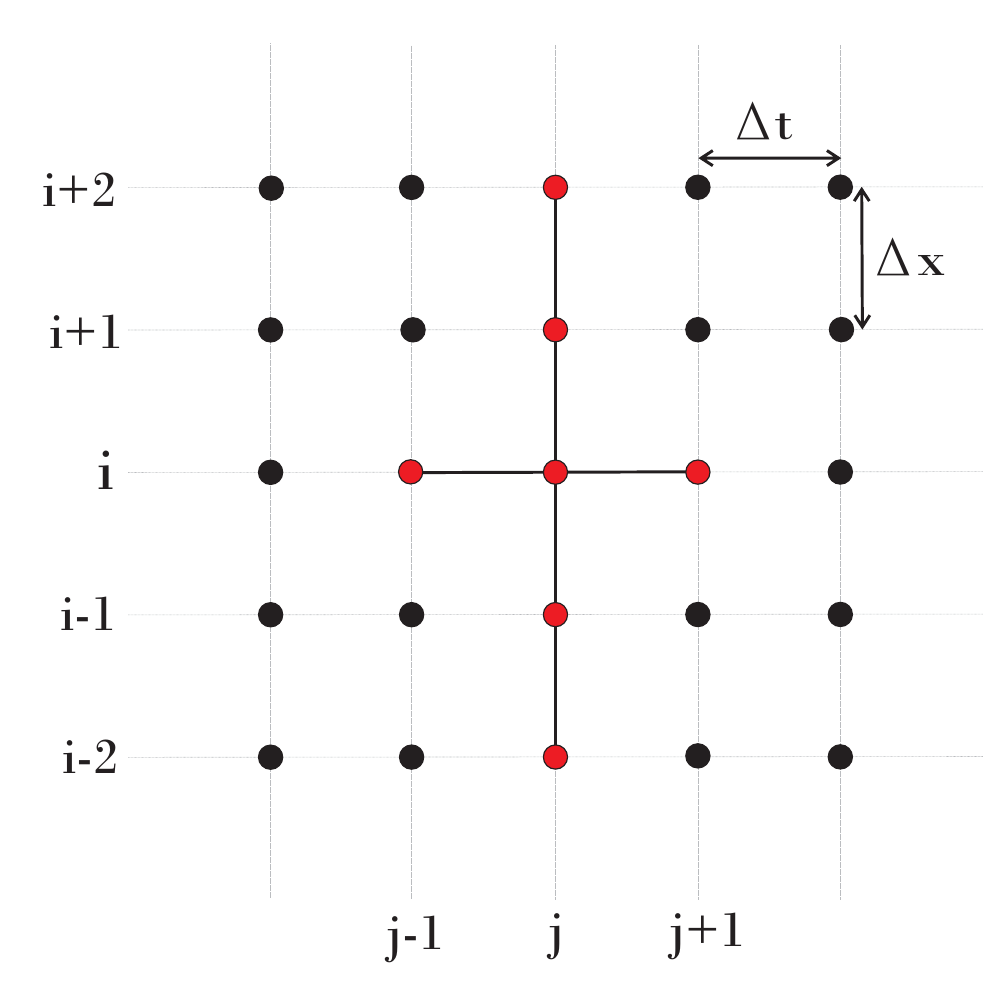}
    \caption{Two-dimensional fourth/second finite-difference discretization, see \cite{molina} for further details. Our numerical results were performed using the python code  \texttt{finite-difference-GM.py} which is publicly available at \cite{finiteElementsGM}. }
    \label{fig:grade}
\end{figure}

\begin{figure}[t]
\includegraphics[width=\linewidth]{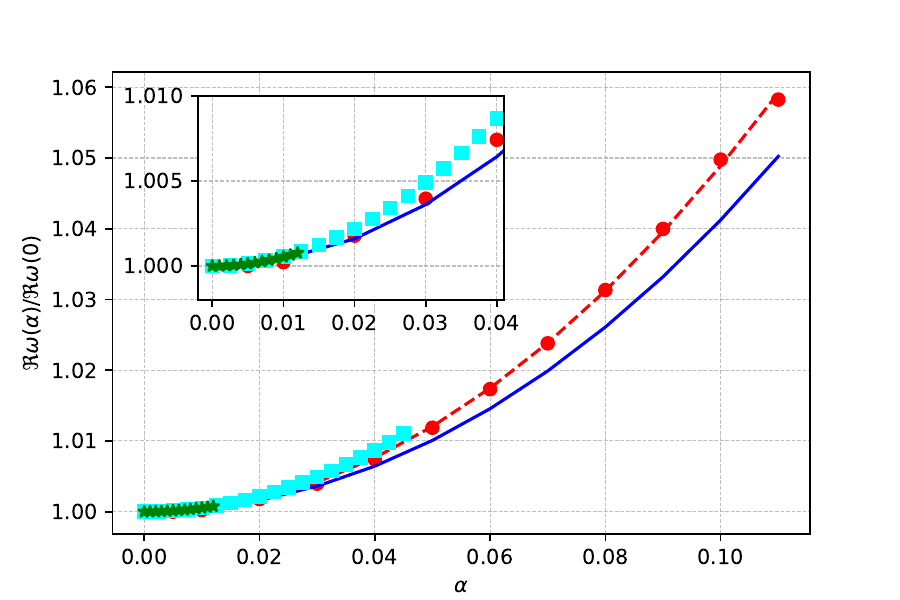}
\includegraphics[width=\linewidth]{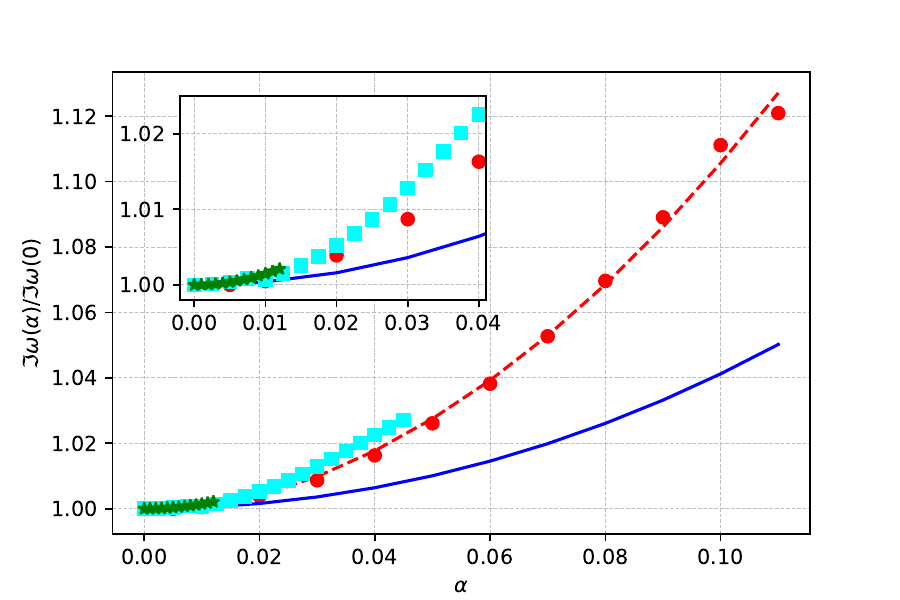}
\caption{
Dependence of the complex fundamental quasinormal frequencies $\omega$ on the parameter $\alpha$. The blue solid line is the variation expected from the dependence of $M^{-1}$ on $\alpha$, namely $\frac{1+\alpha^2}{1-3\alpha^2}$, see (\ref{eq:MAS:GMmass}). The (red) circles,
(cyan) squares, and (green) stars are, respectively, the quasinormal frequencies for $\tilde{x}_{\mathrm{min}}=-5, -10,$ and $-15$. The dashed lines are the quadratic fit $1+b\alpha^2$. The values of $b$ are, respectively, 
$b \approx 4.9,\, 5.4, $ and $5.4$ (Top), and $9.7,\, 13.7, $ and $14.8$ (Bottom).} 
\label{fig:x-freq}
\end{figure}

In order to solve the differential equation~(\ref{eq:MP:ScalWave4}), we employ
the same strategy used in \cite{molina}, namely
a finite-difference scheme  with the spatial derivative calculated at fourth order  and the time derivative at second order,  see Figure \ref{fig:grade}.
To avoid unnecessary clumsy notation, we will drop the $\ell$ and $m$ indices and denote the integration lattice position by
a par of integers $i$ and $j$. 
 At our proposed accuracy level, the second derivatives are
\begin{eqnarray}
    \label{eq:Sim:DiscreteDerivativeX}
  \partial_{\tilde x}^2u_{i,j} &=&\frac{1}{12 (\Delta \tilde x)^2} \left( -u_{i -2, j} + 16 u_{i -1, j} - 30 u_{i , j} \right.\nonumber \\
    & & \quad \quad  \quad \quad  \quad \quad   +\left.  16 u_{i +1, j} - u_{i +2, j} \right)
\end{eqnarray}
and
\begin{equation}
    \label{eq:Sim:DiscreteDerivativeT}
    \partial_t^2 u_{i , j}  = \frac{1}{\Delta t^2} \left(u_{i , j+1}  - 2 u_{i , j}  - u_{i , j-1}  \right).
\end{equation}
The parameter space of our model is spanned  $\alpha$ (constant related to the magnetic field), $x_{\mathrm{min}}$ (position of the mirror), $\ell$ (multipole number of the scalar field) and $n$ (overtone number of the scalar field). This parameter space has been thoroughly explored. For sake of illustration, we present in Table 
\ref{tab:Combined_Frequencies_NoGaps} the results corresponding to three different positions of the mirror: $\tilde{x}_{\mathrm{min}}=-15, -10,$ and $-5$, and
a cutoff $\ell_{\mathrm{max}}=5$. 
For each of these $\tilde{x}_{\mathrm{min}}$ values, the range of $\alpha$ was chosen in equal intervals from zero up to the convergence limit,  
respectively, 
  $\alpha_{\mathrm{max}} = 0.01,\, 0.03,$ and $ 0.11$.
\begin{table}[b]
\caption{Fundamental modes for different mirror positions and cutoff $\ell_{\mathrm{max}}=5$.}
\sffamily
\centering
\arrayrulecolor{white}
\arrayrulewidth=1pt
\renewcommand{\arraystretch}{1.5} 
\resizebox{0.48\textwidth}{!}{
\rowcolors[\hline]{3}{.!50!White}{}
\begin{tabular}{|A|B|C|C|}
\hline
\rowcolor{.!60!Black}
\arraycolor{White}$\mathbf{\alpha}$ & \arraycolor{White}$\tilde{x}_{\mathrm{min}}=-5$ & \arraycolor{White}$\tilde{x}_{\mathrm{min}}=-10$ & \arraycolor{White}$\tilde{x}_{\mathrm{min}}=-15$ \\
\hline
0     & $0.3717 + \mathrm{i}\ 0.0676$  & $0.284242 + \mathrm{i}\ 0.008034$ & $0.211650 + \mathrm{i}\ 0.000749$ \\
0.002 &          &          &                    $0.211654 + \mathrm{i}\ 0.000749$ \\
0.0025 &          &          $0.284253 + \mathrm{i}\ 0.008035$ &   \\
0.005&   $0.3725 + \mathrm{i}\ 0.0692$          & $0.284282 + \mathrm{i}\ 0.008037$  & $0.211678 + \mathrm{i}\ 0.000749$ \\
0.007 &          &          &                     $0.211705 + \mathrm{i}\ 0.000749$  \\
0.0075 &          &           $0.284330 + \mathrm{i}\ 0.008041$ &    \\
0.01 & $0.3725 + \mathrm{i}\ 0.0692$   & $0.284397 + \mathrm{i}\ 0.008039$ & $0.211763 + \mathrm{i}\ 0.000749$ \\
0.015 &          &          $0.284590 + \mathrm{i}\ 0.008054$  &        \\
0.02& $0.3731 + \mathrm{i}\ 0.0694$ & $0.284860  + \mathrm{i}\ 0.008075$ &                \\
0.03 & $0.3739 + \mathrm{i}\ 0.0698$ & $0.285635 + \mathrm{i}\ 0.008137$ &  \\
0.04 & $0.3752 + \mathrm{i}\ 0.0703$ &    &  \\
0.05 & $0.3769 + \mathrm{i}\ 0.0710$  &          &           \\
0.08 & $0.3841+ \mathrm{i}\  0.0740$  &          &            \\
0.11& $0.3942 + \mathrm{i}\ 0.0775$ &  &         \\
\hline
\end{tabular}
}
\label{tab:Combined_Frequencies_NoGaps}
\end{table}
For all considered cases we have found usual decaying quasinormal  oscillations,   for all tested values of $\alpha$. Moreover, we found out that the stronger the magnetic dipole intensity, the larger is 
the imaginary part of the frequency of the QNM.  
The dependence of both the real and imaginary parts of the quasinormal frequencies on $\alpha$ can be well approximated by the quadratic 
expression $1+b\alpha^2$
with $b>0$, see Figure~\ref{fig:x-freq}. Such a quadratic  behavior is not surprising, at least for small $\alpha$, since both metric functions $f$ and $g$ depend explicitly only on the square of $\alpha$. 

\begin{figure}[t]
\includegraphics[width=\linewidth]{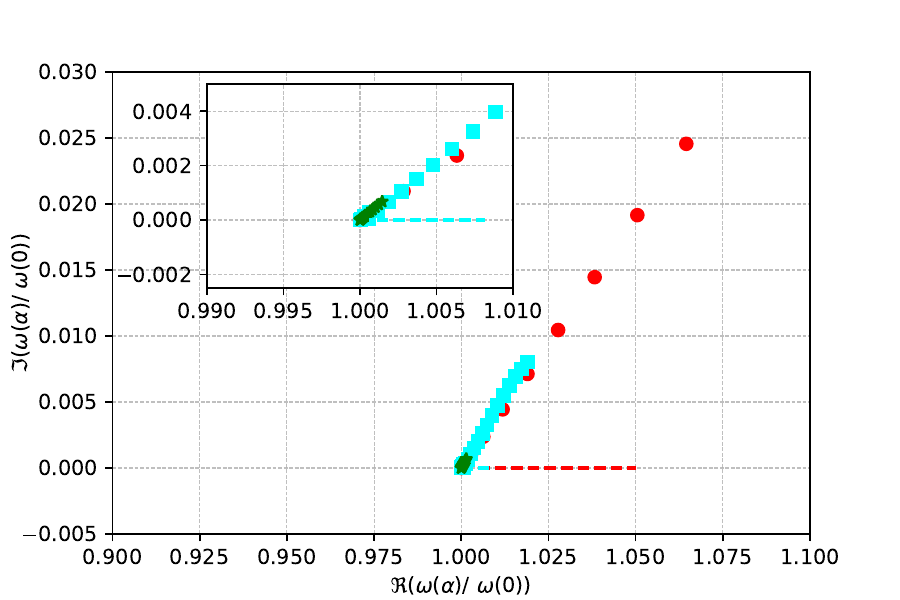}
\caption{Path in the complex plane of the curve given by
$\omega(\alpha)/\omega(0)$.  The (red) circles,
(cyan) squares, and (green) stars correspond, respectively, to $\tilde{x}_{\mathrm{min}}=-5, -10,$ and $-15$. Notice the clear deviation from
the case with a simple dependence of the type $\omega(\alpha)\propto M^{-1}(\alpha)$, which would correspond to the dashed horizontal lines. This a clear indication that the QNM spectrum does depend directly on the magnetization of the compact body, and not only indirectly through  the mass dependence on $\alpha$. } 
\label{fig5}
\end{figure}

It is important to stress that the obtained quadratic dependence on $\alpha$ is different from that one would initially  expected from the dependence of the mass $M$ of the star on $\alpha$ given by (\ref{eq:MAS:GMmass}), and this is indeed one of our main
findings. In analogy with the
usual Schwarzschild and other black hole examples, one might expect the quasinormal frequencies to be proportional to $M^{-1}$ and, hence, proportional to $1+4\alpha^2$ for small $\alpha$. Nevertheless, 
from our fits, we have that the $b$
coefficient for the 
imaginary part of the frequencies is always greater than 
the real part,  meaning that the ``stabilization'' phenomena of the quasinormal modes is not solely due the mass variations, but it indeed increases with the magnetization of the star, see  Figure~\ref{fig5}.
 It is worth noticing that the effect of $\alpha$, and hence of the magnetic field, on the perturbations was typically small, on the order of a hundredth to a few tenths of the value of the frequency, and thus much smaller than the effect of the size of the star.
\begin{figure}[t]
 \includegraphics[width=\linewidth]{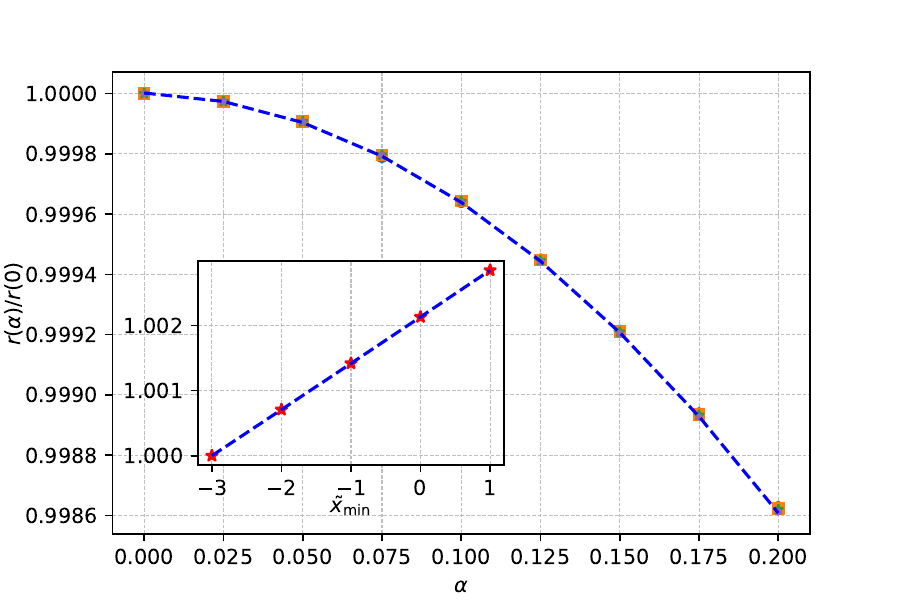}
    \caption{The  power-law tail $t^{-r}$   as function of  
     $\tilde{x}_{\mathrm{min}}$   and 
    $\alpha$. The data is well fitted by the quadratic expression
  $r_0(1+a\tilde{x}_{\mathrm{min}})(1+b\alpha^2)$, see the body text for
  further details. The main graphics depicts the values of
  $r(\alpha)$ for $\tilde{x}_{\mathrm{min}} = -3, -3, -1, 0,$ and $1$. All data points are clustered up to the graphics resolution. The dashed line is the
  quadratic dependence on $\alpha$. The inset shows the linear dependence of
  $r(0)$ on $\tilde{x}_{\mathrm{min}}$.}
    \label{fig:Tail}
\end{figure}
Even for the maximum excursion of $\alpha$ studied in these runs, the real frequency differed from the non-magnetized case by 5\%, while the imaginary frequency varied by about 15\%. However, a variation of the mirror from $\tilde{x}_{\mathrm{min}} = -10$ to $\tilde{x}_{\mathrm{min}} = -5$, corresponding to a variation of 2\% in the radius of the star, would produce a variation of a tenfold factor for the imaginary frequency and of about 30\% for the real frequency.

For all considered cases, the quasinormal oscillations were followed by a final power-law tail $\propto t^{-r}$. Such tails also have their behavior dependent on the magnetization and the size of the star. Figure \ref{fig:Tail} depicts
the results for  $\tilde{x}_{\mathrm{min}}\in [-3,1]$ and the magnetization 
parameter $\alpha\in [0,2]$. These parameters have antagonistic effects on the
power-law tail. The dependence of the power-law exponent $r$ on $\tilde{x}_{\mathrm{min}}$
and $\alpha$ is well fitted by the quadratic expression
$r_0(1+a\tilde{x}_{\mathrm{min}})(1+b\alpha^2)$, with $a\approx 7\times 10^{-4}$,
$b\approx -3.3\times 10^{-2}$, and $r_0\approx 5.97$. 
 As one can see, 
 the power-law exponent $r$  increases linearly with the size of the star, meaning that the tail decays faster, and, conversely, for fixed $\tilde{x}_{\mathrm{min}}$ , it decreases quadratically, implying that the tail decays at a slower pace. Nevertheless, the values of $r$  accumulate  around   $r=6$, the expected   power-law tail of a Schwarzschild Black Hole perturbed by static initial conditions.
 
Finally, it is worth noticing that the position of the mirror boundary condition also affects the ring-down phase of the scalar perturbation. As the surface of the star approaches the peak of the potential, the number of oscillation periods before the onset of the power-law tail also decreases. Conversely, the modes for smaller stars have longer ring-down times. For instance, the results for $\tilde{x}_{\mathrm{min}} = -5$ have few spatial points in which the wave could be analyzed and the ring-down phase reconstructed.

\subsection{The truncated Schwarzschild limit}

The $\alpha=0$ limit of our problem can be solved analytically in  terms of the Heun functions. In this limit, we have
$f=g=1$ and Eq.~(\ref{wave}) simplifies considerably. By using $u_{\ell\omega} = e^{i\omega t}v_{\ell\omega}$ in the 
decomposition (\ref{deco}), we have the following ordinary 
 differential equation 
\begin{eqnarray}\label{xradialEq1}
&& \frac{\mathrm{d}^{2}v }{\mathrm{d}x^{2}}+\frac{2}{x^{2}-1}\frac {\mathrm{d}v }{\mathrm{d}x} \\
&& -\frac{1}{x-1}\left[  \frac{\ell \left( 	\ell+1\right)  }{x+1}-\frac{\left(  x+1\right)  ^{2} \omega^{2}}{x-1}+\frac {2}{\left(  x+1\right)  ^{2}}\right]  v   =0 \, ,\nonumber 
\end{eqnarray}
where, for simplicity, we have dropped the $\ell \omega$ indices. Equation
(\ref{xradialEq1}) 
   can be extended to the entire complex plane and recast into the standard confluent Heun  differential equation   \cite{Heun} 
\begin{equation}
      \frac{\mathrm{d}^{2}v}{\mathrm{d}z^{2}}+\left(  \epsilon+\frac{\beta+1}{z}%
+\frac{\gamma+1}{z-1}\right)  \frac{\mathrm{d}v}{\mathrm{d}z}+\left(
\frac{\mu}{z}+\frac{\nu}{z-1}\right) v=0 ,\label{eqCHeun}
\end{equation}
with
\begin{eqnarray}
\mu &=& \frac{1}{2}(\epsilon -\beta -\gamma + \epsilon\beta -\beta\gamma) -\eta ,\\
       \nu &=& \frac{1}{2}(\epsilon +\beta +\gamma + \epsilon\gamma +\beta\gamma) +\eta .
\end{eqnarray}
The confluent Heun  differential equation appears frequently in QNM studies, see for instance \cite{Heun, fiziev2006exact,Horta_su_2018,bruno1,bruno2}. Their
solutions  have two regular singular points ($z=0,1$) and one irregular point located at infinity. Local solutions of Eq.~(\ref{eqCHeun}) around the regular singular points are written in terms of the Heun confluent function, 
 often denoted by
$\mathrm{HeunC}\left( \epsilon,\beta,\gamma,\delta \label{HeunEq}
,\eta;z\right)$.
Solutions of Eq.~(\ref{eqCHeun}) around the point $z=0$ can be written as  a linear combination  of the functions
\begin{subequations}
\begin{align}
H_0^{+}\left( z\right) &=\mathrm{HeunC}\left( \epsilon,\beta,\gamma,\delta,\eta;z\right)  , \label{H0+} \\
H_0^{-}\left( z\right) &=z^{-\beta}\mathrm{HeunC}\left( \epsilon,-\beta,\gamma,\delta,\eta;z\right) , \label{H0-}
\end{align}
\end{subequations}
For $z \to \infty$, it is possible to derive  asymptotic series \cite{Heun} 
such that the linearly independent solutions take the forms 
\begin{subequations}
\begin{align}
H_{\infty}^{+}\left(  z\right)    & \sim\frac{e^{-\beta
\left(  z+\ln z\right)  }}{z}\sum_{\nu\geq0}\frac{a_{\nu}^{+}}{z^{\nu}}\; , \text{
\ \ }a_{0}^{+}=1  ,\\
H_{\infty}^{  - }\left(  z\right)    & \sim\frac{1}{z}\sum
_{\nu\geq0}\frac{a_{\nu}^{-}}{z^{\nu}}\; , \text{ \ \ }a_{0}^{-}=1  ,
\end{align}
\end{subequations}
with the coefficients $a_{\nu}^{\pm}$ determined by the known recurrence relations of the Heun functions \cite{Heun}.
Thus, the general solution $U_{X}(z)$ around the point $X=0$ or $X=\infty$  can be written as
\begin{align}
U_X(z) = C^{+}_{X}H^{+}_{X}\left(  z\right)+C^{-}_{X}H^{-}_{X}\left(  z\right) ,
\end{align}
where $C^{\pm}_X$ are constants.

Let us  return to the original problem of solving Eq.~(\ref{xradialEq1}) with the Dirichlet boundary condition on the surface of the star,
\begin{equation}
v(x_{\mathrm{min}})=0   .\label{DBC}
\end{equation}
It is convenient here to introduce the 
the following ansatz
\begin{equation}
v \left(  z\right)  =(1-z)^{2i\omega}\left(  z+1\right)e^{i\omega z} 
w  .\label{ansatz}
\end{equation}%
One can see that the function $w$ also satisfies a confluent Heun differential equation. 
In terms of this ansatz, the solution of the confluent Heun equation  (\ref{eqCHeun}) obeying the Dirichlet
boundary condition (\ref{DBC}) reads 
\begin{widetext}
\begin{align}\label{sol1}
&v (z) = C_{1}(1-z)^{2ik\omega} (z+1) e^{ik\omega z}\times \\
& \Bigg[\mathrm{HeunC}\left(\epsilon,\beta,0,\delta,\eta;\frac{1-z}{2}\right)  
-  \left(\frac{1-x_{\mathrm{min}}}{1-z}\right)^{4ik\omega}\!\!\!
\frac{\mathrm{HeunC}\left(\epsilon,\beta,0,\delta,\eta;\frac{1-x_{\mathrm{min}}}{2}\right)}{\mathrm{HeunC}\left(\epsilon,-\beta,0,\delta,\eta;\frac{1-x_{\mathrm{min}}}{2}\right)} 
 \mathrm{HeunC}\left(\epsilon,-\beta,0,\delta,\eta;\frac{1-z}{2}\right)\Bigg] ,\nonumber
\end{align}
\end{widetext}
where
\begin{subequations}
\begin{align}
\epsilon &  =-4i\omega, \quad
\beta   =4i\omega, \quad \gamma   =0,\\
\delta &  =-8\omega^{2}, \quad
\eta   =8\omega^{2}- \ell \left(  \ell+1\right).
\end{align}
\end{subequations}

The quasi-normal spectrum can be directly extracted from the solution~(\ref{sol1}). 
To accomplish this, we must ensure that there are no ingoing modes coming from infinity. The solution can be also written in terms of asymptotic series at infinity,
\begin{equation}
    v (z)=C_{\infty}^{+}U^{+}_{\infty}(z)+C_{\infty}^{-}U^{-}_{\infty}(z),\label{infinitosolution}
\end{equation}
where, in accordance with our ansatz (\ref{ansatz}),
\begin{equation}
U_\infty^{\pm}(z)=(1-z)^{2i\omega}(z+1)e^{i\omega z}H_\infty^{\pm}(z) .
\end{equation}  
 {The next step is} to find the condition that implies   $C_{\infty}^{-}=0$, since this is the requirement that no wave should come from infinity. By exploiting the fact that the solution approaches zero as $z\to \infty$, it is possible to constraint the constants  
$ C^{\pm}_{\infty} $ as
\begin{equation}
    \left | C_{\infty}^{-} \right |= \left | C_{\infty}^{+} \right |\lim_{z\to \infty}\left|\frac{U^{+}_{\infty}(z)}{U^{-}_{\infty}(z)}\right|, \label{C-}
\end{equation}
from where we have that $C_{\infty}^{-} = 0$ for
\begin{equation}
    \lim_{z\to \infty}\left|U^{-}_{\infty}(z)\right| = \infty .
    \label{condition}
\end{equation}
As noted in \cite{fiziev2006exact}, condition~(\ref{condition}) is satisfied  in the lower half of the complex plane, 
\begin{equation}
\mathrm{\arg}\left(  z\right)  +\mathrm{\arg}\left(\omega\right)  \in\left(  -\pi,0\right)   .
\end{equation}
with optimal convergence  for $\mathrm{arg}(z)= -\pi/2-\mathrm{arg}(\omega)$.  
Hence, we have finally  a transcendental equation in terms of the confluent Heun functions,
\begin{widetext}
\begin{gather}
    \lim_{_{\substack{z \rightarrow\infty\\\mathrm{\arg}\left(  z\right)  =-\frac{\pi}{2}-\mathrm{\arg}\left(\omega\right)  }}}\Bigg| \left(  1-z\right)  ^{4i\omega}\frac {\mathrm{HeunC}\left(  \epsilon,\beta,0,\delta,\eta;\frac{1-z}{2}\right) }{\mathrm{HeunC}\left(  \epsilon,-\beta,0,\delta,\eta;\frac{1-z}{2}\right) }  
-\left(  1-x_{\mathrm{min}}\right)  ^{4i\omega}\frac{\mathrm{HeunC}\left( \epsilon,\beta,0,\delta,\eta;\frac{1-x_{\mathrm{min}}}{2}\right)  }{\mathrm{HeunC}\left(  \epsilon,-\beta,0,\delta,\eta;\frac{1-x_{\mathrm{min}}}{2}\right)  }\Bigg |=0 \, . 
\label{eqtransc}
\end{gather}
\end{widetext}
  Eq.~(\ref{eqtransc}) is valid only for discrete values of $\omega_n$,  with the first mode, corresponding to $n=0$,   referred to as the fundamental mode.
   Actually, there are several approaches to solve the zero-finding 
   problem for equations like 
   (\ref{eqtransc}). We have opted to recast the complex   equation 
   in  a system of two real equations in two real variables,   and use a standard Newton  method to numerically determine the first modes, whose values are shown in Table \ref{tab:Modes_Frequencies}. As one can see, we have a good agreement with the numerical results of Table \ref{tab:Combined_Frequencies_NoGaps}, indicating that our numerical and analytical approaches are indeed robust.
\begin{table}[t]
\caption{The first four modes for $\ell=1$ for different values of $\tilde{x}_\mathrm{min}$. These values where determined numerically by
adopting $|z_\infty|=100$, which was enough to guarantee a good convergence of the results.  }
\centering
\sffamily
\arrayrulecolor{white}
\arrayrulewidth=1pt
\renewcommand{\arraystretch}{1.5} 
\resizebox{0.48\textwidth}{!}{
\rowcolors[\hline]{3}{.!50!White}{}
\begin{tabular}{|A|B|C|C|} 
\hline
\rowcolor{.!60!Black}
\arraycolor{White}\textbf{n} & \arraycolor{White}$\tilde{x}_\mathrm{min}=-5$ & \arraycolor{White}$\tilde{x}_\mathrm{min}=-10$ & \arraycolor{White}$\tilde{x}_\mathrm{min}=-15$ \\
\hline
0 & $0.37033 + \mathrm{i}\ 0.06688$ & $0.28409 + \mathrm{i}\ 0.00802$ & $0.21155 + \mathrm{i}\ 0.00074$ \\
1 & $0.54250 + \mathrm{i}\ 0.32595$ & $0.45383 + \mathrm{i}\ 0.10192$ & $0.36120 + \mathrm{i}\ 0.02769$ \\
2 & $0.75345 + \mathrm{i}\ 0.62517$ & $0.64775 + \mathrm{i}\ 0.24144$ & $0.50976 + \mathrm{i}\ 0.09814$ \\
3 & $0.98312 + \mathrm{i}\ 0.92643$ & $0.85580 + \mathrm{i}\ 0.38376$ & $0.67266 + \mathrm{i}\ 0.17503$ \\
\hline
\end{tabular}
}
\label{tab:Modes_Frequencies}
\end{table}

\section{Final remarks}

We have shown that the spheroidal magnetized compact body described 
by a simple modification  of the Gutsunaev-Manko  is well-defined and stable. Moreover, we have observed that the magnetic field has a slightly stabilizing effect on QNM spectrum, {\it i.e.},  greater magnetization leads to faster decay of the modes. An opposite effect was noted in the tails regime.

For any practical purpose, we must be able to express the mass $M$ and 
 the magnetic moment $\mu$ of our compact object in terms of usual astrophysical units. This can be easily done by recasting the $G_N$ and $c$ factors in (\ref{eq:MAS:GMmass}) and (\ref{eq:MAS:GMdipole}). Recalling that $G_NM_\odot /c^2 \approx 1.48$ km, we have 
\begin{equation}
   M  =\frac{1-3\alpha^2}{1+\alpha^2}\left(\frac{k}{1.48 \,{\rm km}}\right)M_\odot,\label{physicalMass} 
\end{equation}
with, of course, $k$ given in km. For the magnetic moment $\mu$, the situation is
a little bit more intricate. From (\ref{eq:MAS:Maglim2}), we have that the magnitude
of the magnetic field in our solution is such that $|B|\propto \mu/r^3$. Hence, to 
accommodate the typical magnetic field ($10^{13}$ to $10^{15}$ G) of a magnetar with superficial radius of $10$ km, the magnetic moment $\mu$ must be of the order
$10^{16}$ to $10^{18} \,{\rm G}\,{\rm km}^3$.
Recalling that in Gaussian units the magnetic dipole moment $\mu$ is expressed in ${\rm erg} / {\rm G}= {\rm cm}^\frac{5}{2} {\rm g}^\frac{1}{2}{\rm s}^{-1}$, 
we have that  
$\sqrt{G}\mu/{c^2}$ has units ${\rm cm}^2$ and the units of (\ref{eq:MAS:GMdipole}) can be restored straightforwardly. 
\begin{equation}
      \mu  =\frac{8\alpha^3}{\left(1+\alpha^2\right)^2}\frac{c^2k^2}{\sqrt{G_N }}
         =\frac{8\alpha^3}{\left(1+\alpha^2\right)^2} \left(\frac{k}{1.48 \,{\rm km}}\right)^2 \frac{G_N^\frac{3}{2}M^2_\odot}{c^2}
      ,\label{physicaldipole}
\end{equation}
with  
\begin{equation}
  \frac{G_N^\frac{3}{2}M^2_\odot}{c^2} \approx 7.6\times 10^{19} \,{\rm G}\,{\rm km}^3
      .\label{physicaldipole1}
\end{equation}
One can see that our model can describe the typical magnetic field observed in magnetars for small values of $\alpha$.

Although we have focused only on fundamental frequencies for the perfect reflection (mirror) model, the excited modes can also be extracted using   numerical methods such as those ones discussed in \cite{London_2014}.  
Finally, the analysis conducted in this work 
is restricted to uncharged scalar perturbation. Therefore, its straightforward extension would be to introduce other fields with distinct spin and charges.
Electrically charged fields, in particular, would be especially  interesting  since they could interact electromagnetically with the magnetic moment present in our model based on the Gutsunaev-Manko solution.

\section*{Acknowledgments}

The authors want to acknowledge the funding support provided by the following
agencies:
Fapesp, grants 2014/07885-0 (EA), 2021/09293-7 (AS), and 2022/07534-0 (CM); CNPq, 
grants 303592/2020-6 (EA), 141276/2021-5 (MRRJ),  306785/2022-6 (AS), 141310/2024-3 (ECR),
  and 310533/2022-8 (ARQ); CAPES (LF); FINEP (EA); and  FAPESQ-PB (ARQ).
AS also wishes to thank  Vitor Cardoso and José S. Lemos
for the warm hospitality at the Center for Astrophysics and Gravitation of the University of
Lisbon, where this work was finished.

\end{document}